\documentclass[pdflatex,sn-nature]{sn-jnl}% 

\usepackage{graphicx}%
\usepackage{multirow}%
\usepackage{amsmath,amssymb,amsfonts}%
\usepackage{amsthm}%
\usepackage{mathrsfs}%
\usepackage[title]{appendix}%
\usepackage{xcolor}%
\usepackage{textcomp}%
\usepackage{manyfoot}%
\usepackage{booktabs}%
\usepackage{algorithm}%
\usepackage{algorithmicx}%
\usepackage{algpseudocode}%
\usepackage{listings}%
\usepackage{siunitx}%
\DeclareSIUnit{\grav}{g}

\usepackage{pdfpages}
\usepackage{pax}
\theoremstyle{thmstyleone}%
\theoremstyle{thmstyletwo}%

\theoremstyle{thmstylethree}%

\begin{document}

%\title[Article Title]{Macroscopic free mass under continuous quantum measurement}
%\title[Article Title]{\makebox[\textwidth][c]{Macroscopic free mass under}\\[-2.0ex]\makebox[\textwidth][c]{continuous quantum measurement}}
\title[Article Title]{\makebox[\textwidth][c]{Taming a macroscopic free mass through}\\[-0.0ex]\makebox[\textwidth][c]{continuous quantum measurement}}

%%=============================================================%%
%% GivenName	-> \fnm{Joergen W.}
%% Particle	-> \spfx{van der} -> surname prefix
%% FamilyName	-> \sur{Ploeg}
%% Suffix	-> \sfx{IV}
%% \author*[1,2]{\fnm{Joergen W.} \spfx{van der} \sur{Ploeg} 
%%  \sfx{IV}}\email{iauthor@gmail.com}
%%=============================================================%%

\author[1]{\fnm{Umang} \sur{Mishra}}
\equalcont{These authors contributed equally to this work.}

\author[1]{\fnm{Teodor} \sur{Strömberg}}
\equalcont{These authors contributed equally to this work.}

\author[1]{\fnm{Gabriela} \sur{Valencia-Guzman}}

\author[1]{\fnm{Felipe} \sur{Gómez-Lozada}}

\author[1]{\fnm{Marta} \sur{Marchese}}

\author*[1]{\fnm{Onur} \sur{Hosten}}\email{onur.hosten@ist.ac.at}

\affil[1]{\orgname{Institute of Science and Technology Austria}, \orgaddress{\city{Klosterneuburg}, \postcode{3400}, \country{Austria}}}

%\affil[2]{\orgdiv{Department}, \orgname{Organization}, \orgaddress{\street{Street}, \city{City}, \postcode{10587}, \state{State}, \country{Country}}}

%\affil[3]{\orgdiv{Department}, \orgname{Organization}, \orgaddress{\street{Street}, \city{City}, \postcode{610101}, \state{State}, \country{Country}}}

%%================================%%
%% Sample for structured abstract %%
%%================================%%

\abstract{The free particle is the most elementary dynamical system in quantum mechanics. Despite remarkable progress in quantum control of increasingly massive solid-state mechanical oscillators, the analogous control of the free-particle counterpart has remained elusive. Dissipation and environmental noise have prevented such control in the inherently low-frequency domain of free-particle dynamics---a regime of growing importance for quantum sensing and tests of fundamental physics. Here we realize a suspended 1-milligram rotor operating in the free-particle regime with exceptional environmental isolation. Combining this isolation with a low-noise measurement architecture, we directly resolve thermal fluctuations in the motion associated with a damping rate of only 7~nHz. Through continuous quantum measurement of the rotor, we localize its otherwise unconfined motion, obtaining conditional states with entropic temperatures as low as 140~nK as a quantitative measure of control in the free-particle regime. This is two orders of magnitude below temperatures previously achieved in tabletop optomechanical experiments. We demonstrate immediate utilities of the new operating regime, setting new constraints on modifications of quantum mechanics and benchmarking inferred entanglement-to-decoherence rate ratios nearly two orders of magnitude beyond the state of the art for tests of gravity at the quantum-classical interface.}

%\keywords{keyword1, Keyword2, Keyword3, Keyword4}

%%\pacs[JEL Classification]{D8, H51}

%%\pacs[MSC Classification]{35A01, 65L10, 65L12, 65L20, 65L70}

%\makeatletter\lineno@off\makeatother
\maketitle
%\clearpage
%\linenoon

\section{Introduction}\label{sec:Introduction}

The ability to prepare and control mechanical systems near their quantum limits has advanced dramatically over the past two decades. Mechanical devices spanning more than 20 orders of magnitude in mass---from femtogram nanomechanical resonators to kilogram-scale gravitational-wave detectors---have become central to quantum science and precision sensing~\cite{delic2020cooling,rossi2018measurement,bild2023schrodinger,whittle2021approaching,underwood2015measurement}. These advances, however, have almost exclusively concerned harmonic oscillators rather than free masses, with resonance frequencies typically ranging from sub-kilohertz to gigahertz.

This distinction is becoming increasingly important. Many emerging directions in quantum science and fundamental physics---including testing the quantum nature of gravity~\cite{bose2017spin,marletto2017gravitationally,oppenheim2023postquantum,dorau2026quantum}, proposed modifications of quantum mechanics, dark-matter models and new short-range forces~\cite{carlesso2022present,bose2025massive}---require mechanical systems combining large mass, long coherence times and quantum-limited control. Perhaps counterintuitively, low-frequency mechanical systems at directly visible macroscopic scales are particularly well suited to meeting many of these requirements, with progressively weaker confinement ultimately approaching the free-particle regime. In this regime, as has been central to proposals both for quantum state control and sensing~\cite{weiss2021large,rossi2025quantum,krisnanda2020observable}, extended intervals of free expansion help engineer interesting quantum states or enhance sensitivity to physical phenomena, while trapping primarily serves to initialize experiments.

To date, experimental developments have largely followed two separate directions. Free-particle dynamics has been exploited for precision force sensing, most notably in gravitational-wave detectors~\cite{abbott2016observation, danilishin2012quantum}, whereas demonstrations of low-entropy states have relied on trapped oscillators, including the electrostatically confined mirrors of gravitational-wave detectors~\cite{whittle2021approaching}. As we show, the combination of these two directions---exploring and exploiting conditional states of a macroscopic free particle under continuous quantum measurement---enables capabilities and performance beyond those demonstrated with trapped systems. Central to this advance is continuous quantum measurement, which provides an alternative to confinement for accessing localized conditional states, as recently envisioned in proposals for generating nonclassical states of macroscopic objects~\cite{muller2008entanglement,meng2020mechanical,miki2023generating,papa2026remote}.

Here we realize a suspended milligram-scale rotor operating in the free-particle regime under continuous quantum measurement, establish a framework for understanding the resulting conditional states, and demonstrate immediate utilities in fundamental science.

\begin{figure}[h]
\centering
\includegraphics[width=0.5\textwidth]{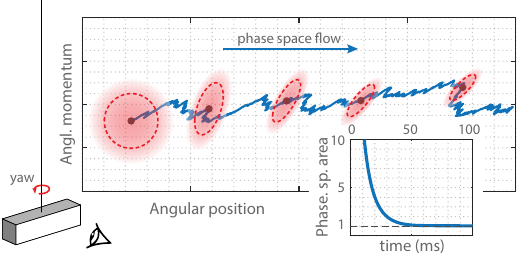}
\caption{Conditional states of a free mass. Continuous measurements infer a trajectory for the centroid of the rotor’s yaw motion, tracking it in phase space with an inference uncertainty progressively shrinking to a steady state. The uncertainty ellipse shape is determined by the momentum-position correlating dynamics of the free mass. Inset: Illustration of the timescales over which the conditional phase-space area approaches its steady-state value---plotted relative to the steady-state value.}\label{fig:Concepts}
\end{figure}

\section{Concepts}\label{sec:Concepts}
The physics of a continuously measured free particle is best understood on its own terms, rather than simply as the zero-frequency limit of a harmonic oscillator. The latter perspective can obscure a central feature of the system: continuous measurement introduces a new characteristic frequency scale, even in the absence of a restoring force. We therefore develop the concepts underlying this work with minimal reliance on the conventional framework of harmonic oscillators, drawing comparisons with them where they provide useful context.

A free-particle in isolation indeed does not have a frequency or length scale associated with it. However, continuous measurement fundamentally changes this by localizing the particle's state in phase space, with measurements in a sense substituting the trapping familiar from oscillators: A continuously monitored free particle evolves towards a steady-state conditional Gaussian distribution in phase space, having a finite stationary entropy. Physically, while thermal/quantum fluctuations lead to diffusion in phase space, measurements counteract to reveal the trajectory (Fig.~\ref{fig:Concepts}). This purifies the state---in a manner analogous to those demonstrated with oscillators ~\cite{wieczorek2015optimal,rossi2019observing}---with the residual uncertainty about the inferred phase-space centroid constituting the remaining entropy.

\begin{figure}[h]
\centering
\includegraphics[width=0.55\textwidth]{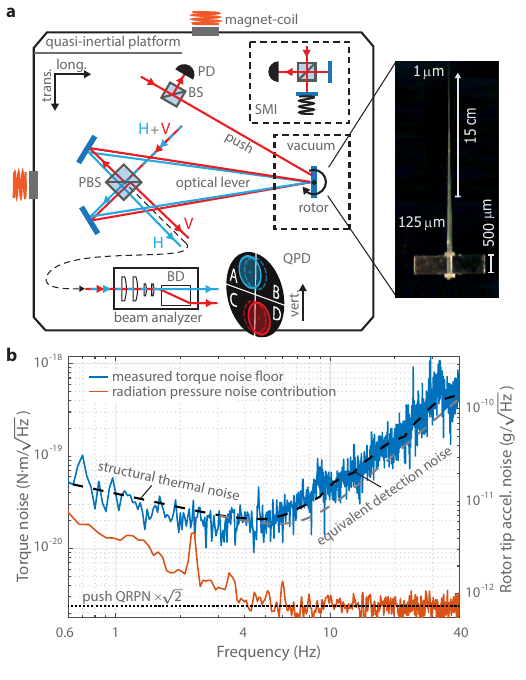}
\caption{Experimental setup. \textbf{a.} The dielectric-mirror-coated $0.5\times0.5\times\SI{2}{\milli\meter^3}$ fused-silica rotor is suspended in an ultrahigh-vacuum chamber. Its yaw motion is measured using a `double' optical lever that counter-circulates a pair of horizontal (H) and vertical (V) polarized beams, whose displacements are observed differentially on a quadrant photodiode (QPD)---operation A-B-C+D. The yaw motion is manipulated with radiation pressure through the ‘push’ beam, in a feedback loop with the optical lever, for damping, aligning or trapping the rotor. The whole setup rests on a passive seismic isolation platform acting as an inertial mass above its resonance frequency of 0.4~Hz in all directions. Magnet-coil pairs actuate motion of the platform, in a feedback loop with the optical lever, to continuously damp higher order mechanical modes. Residual environmental disturbances on the platform are sensed with a suspended-mirror Michelson interferometer (SMI) for feed-forward compensation of angle measurements. All beams are delivered from a single DFB laser operating at \SI{780}{\nano\meter}: \SI{40}{\micro\watt} lever, \SI{1.5}{\milli\watt} push and \SI{100}{\micro\watt} SMI. BD: calcite beam displacer, BS: beam splitter, PBS: polarizing BS, PD: photodiode. \textbf{b.} Near-thermal-torque-noise limited operation---2 min. continuous data taken under trapped condition at \SI{240}{\milli\hertz}---verifying the expected ultralow dissipation rate around the minimal noise region of \SI{6}{\hertz}. The combined contributions of ideal thermal torque noise and measured detection-imprecision noise are overlaid for comparison (black-dashed), with the ideal shot-noise-limited-detection additionally indicated (gray-dashed). Torque noise contribution from the power stabilized push beam is subdominant, lying near its quantum radiation pressure noise (QRPN) limit.}\label{fig:Experiment}
\end{figure}

The available reference frequency $\Omega_\text{ref}$ is now the inverse of the time scale \mbox{$t_\text{ref}=1/\Omega_\text{ref}$} for measurement-conditioned states to reach steady variances in phase space (Fig.~\ref{fig:Concepts}). Incidentally, this frequency is the crossover frequency where the spectrum of position noise falls below the measurement imprecision noise level (Fig.~\ref{fig:Experiment}b)---in our case, $\Omega_\text{ref}/2\pi=\SI{6}{Hz}$ and $t_\text{ref}=\SI{26}{ms}$. As we show here, the well-defined conditional entropy now also allows an entropic temperature to be assigned to the conditional states through the canonical definition $T=(k_B\tfrac{\partial S}{\partial E})^{-1}$ of temperature---relating the change in entropy $S$ to that in energy $E$, where $k_B$ is Boltzmann's constant. Noting that this is the exact quantitative way that a temperature can be assigned to a single oscillator coupled to an environment~\cite{ali2020quantum}, it allows for an unambiguous comparison of free particles with oscillators. 

Quantum purity $\mathcal{P}~(\leq1)$, which specifies the occupied phase-space area $\mathcal{A}=\pi\mathcal{P}^{-1}$ in units of the reduced Planck constant $\hbar$, facilitates the rigorous connection between continuous quantum measurement and temperature through the von Neumann entropy $S=\mu_+\ln{\mu_+}-\mu_-\ln{\mu_-}$ for Gaussian states, where $\mu_\pm=(\mathcal{P}^{-1}\pm1)/2$. In one dimension, this construction yields (see SI)
\begin{equation}
k_BT=\tfrac{3f(\mathcal{P})}{2}E,\label{Eq:temperature}
\end{equation}
where, $E$ is the kinetic energy in the reference frame of the inferred position centroid, and \mbox{$f(\mathcal{P})=2\mathcal{P}/\text{ln}\tfrac{1+\mathcal{P}}{1-\mathcal{P}}\approx1$} with the approximation valid for \mbox{$\mathcal{P}<0.5$}. Eq.~\ref{Eq:temperature} is independent of any characteristic length or frequency scale. Interestingly, it predicts a minimum achievable temperature of \SI{15}{\nano\kelvin} given our rotor's environmental isolation level (see SI)---pointing at the fundamental non-measurability of a zero temperature using position measurements, with free particles or oscillators alike. This arises from a trade-off between measurement imprecision and back-action, similar to notions of force sensing at the standard quantum limit~\cite{caves1981quantum}: Finer position measurements, per Heisenberg uncertainty principle, increase momentum uncertainty.

Assessing the introduced entropic temperatures requires verifying the measurement-conditioned phase-space distributions. In optomechanical systems, such verification methods have recently been demonstrated, leveraging quantum-state retrodiction~\cite{gammelmark2013past,lammers2024quantum,zhang2017prediction}. While successful for nearly isotropic phase-space distributions~\cite{rossi2019observing,meng2022measurement}, the predictive power of the basic method has been questioned for squeezed phase-space distributions~\cite{sakai2026unbiased}. To clearly characterize the intrinsically squeezed conditional states of a free-particle, here we introduce a novel measurement-based strategy that uses smoothed states~\cite{mayne1966solution,fraser1969optimum,guevara2015quantum,laverick2019quantum} as the main reference.

%While successful for nearly isotropic phase-space distributions~\cite{rossi2019observing,meng2022measurement}, the predictive power of the basic method is lost for squeezed phase-space distributions, requiring modifications in estimation strategies~\cite{sakai2026unbiased}. Since free-particle conditional dynamics is intrinsically squeezed due to position-momentum correlations, we use a novel strategy we use a refined analysis strategy in this work.

\section{Experimental Realization}\label{sec:Experimental}
In our setup, the physical system is a rotor formed by a 1-mg mirror bar, suspended with a \SI{15}{\centi\meter}-long fused silica fiber of \SI{1}{\micro\meter} diameter (Fig.~\ref{fig:Experiment}a). It is housed in a vacuum environment below \SI{1e-10}{\milli\bar}. While technically constituting a \SI{15}{\second} period (\SI{65}{\milli\hertz}) torsional pendulum, in practice it operates deeply in the free particle regime given the $t_\text{ref}=\SI{26}{\milli\second}$ reference timescale of interest in connection with measurement-based state conditioning (Fig.~\ref{fig:Concepts}). The damping of the rotor follows the structural dissipation model~\cite{saulson1990thermal}, in which dissipation is a multi-time-scale process, leading to a timescale-dependent friction strength. Through the fluctuation-dissipation theorem, this leads to thermal torque noise with a power spectral density (PSD) $S_{\tau\tau}=4k_B T_0 I \gamma(\omega)$. Here, the rotor's moment of inertia $I=\SI{3.3e-13}{\kilo\gram\cdot\meter^2}$, and the energy damping rate $\gamma(\omega)=\gamma_\text{ref}\Omega_\text{ref}/\omega$ falls inversely with angular frequency $\omega$. The anticipated damping rate $\gamma_\text{ref}$ at the reference angular frequency $\Omega_\text{ref}$ can be estimated from the damping behavior of the \SI{65}{\milli\hertz} torsional oscillations (see SI). However, for quantum control of mechanical motion, the relevant quantity setting the decoherence rates is not the oscillation decay itself but rather the noise arising from it. Accessing this noise directly has not been possible in the leading ultralow loss platforms~\cite{dania2024ultrahigh,chang2010cavity} which have shown dissipation rates as low as $2\pi\times\SI{69}{\nano\hertz}$. In our system, we can access ultralow torque noise levels near \SI{1.8e-20}{\newton\cdot\meter\per\sqrt{\hertz}}, verifying that there is no excess noise to suggest a dissipation level higher than $\gamma_\text{ref}=2\pi\times\SI{7.3}{\nano\hertz}$---the expected dissipation at our reference timescale---once the thermal and detection noises are accounted for (Fig.~\ref{fig:Experiment}b). The implied acceleration sensitivity of \SI{5.6e-12}{\grav\per\sqrt{\hertz}} for the tip of the mirror bar is nearly two orders of magnitude better than state of the art experiments at the same mass scale~\cite{fuchs2024measuring}.

The detection and control principles (Fig.~\ref{fig:Experiment}a) of the experiment build on those reported in ref.~\cite{agafonova2026one}. The main differences are the utilization of 1) a $10^4$-fold reduced torsional suspension stiffness to reach the desired low thermal noise levels, 2) a novel `double' optical lever for detection noise immunity, 3) a 0.4 Hz cut-off passively isolated platform for motion control, and 4) an auxiliary suspended-mirror Michelson interferometer for platform motion sensing.

The UV-glue-tethered fused silica suspension fiber yields a free running torsional quality factor of $Q_0=102,000$---an order of magnitude improvement over reported laser welded monolithic designs~\cite{catano2020high}, operating five times beyond the expected surface-to-volume-ratio-dependent material limits~\cite{gretarsson1999dissipation}. While $Q_0$ refers to motion at \SI{65}{\milli\hertz}, the reference quality factor commensurate with the observed low noise is $Q_\text{ref}=\Omega_\text{ref}/\gamma_\text{ref}\approx1\times 10^9$---within the range of highest quality factors reported in optomechanical systems~\cite{dania2024ultrahigh,beccari2022strained,rossi2018measurement}. The `double' optical lever renders the detection of the rotor angle insensitive to input beam fluctuations by translating the yaw motion into two counter-tilting beams, allowing near-shot-noise-limited operation down to 1 Hz. The passive seismic isolation platform is used to damp and control the resonant swing, pitch, roll, and violin modes of the pendulum through its actuated motion using magnet-coil pairs. Lastly, the in-air Michelson interferometer is used to establish an independent inertial reference frame by suspending one of its mirrors to extract the absolute residual motion of the platform and to feed-forward compensate the relative platform-rotor angle measurements. For more details on the technical aspects of the setup, see SI. 

\section{State conditioning and characterization}\label{sec:State}

Each experimental run begins by aligning the rotor using a push beam via radiation pressure (Fig.~\ref{fig:Experiment}a), bringing the reflected lever beam within the detection range. The rotor is then actively trapped at \SI{240}{\milli\hertz} at the center of the linear detection range under feedback control using push-beam actuation (see SI). To investigate measurement-conditioned dynamics, we release the rotor from the induced trap by locking the push-beam power at its pre-release mean value and then monitor the angular evolution of the rotor.

\begin{figure}[h]
\centering
\includegraphics[width=1.0\textwidth]{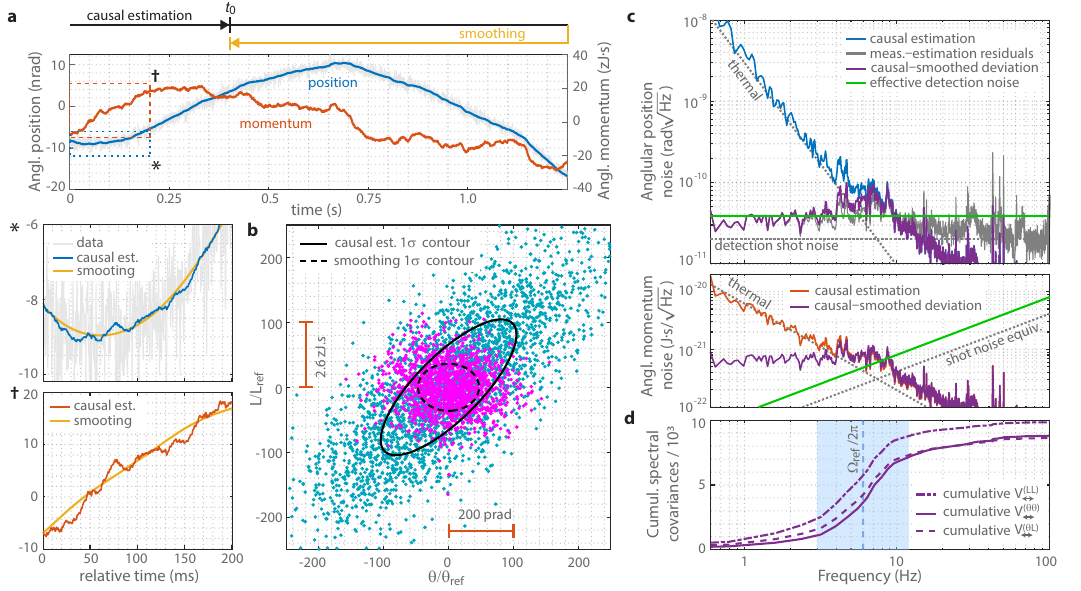}
\caption{Conditional state at a time $t_0$ through causal and smoothed estimations, analyzing 100 consecutive 1.25-s measurement sections obtained with $\SI{320}{\mu s}$ sampling time (see SI for data analysis). \textbf{a.} Example section, showing the rotor's tracked angular position/momentum. A zoom-in to the asterisk/dagger region shows the relation between measurements, causal estimations and smoothed estimations. \textbf{b.} Characterized conditional state distributions for states conditioned through causal and smoothed estimations. The distributions are obtained by streaming a morphed version $\tilde{R}_\leftrightarrow$ (see text) of the causal-smoothed deviation vector---displaying only every $60^{\text{th}}$ sampled element to mitigate visual overcrowding. The axes use the reference scales $\theta_\text{ref}\equiv(2I\Omega_\text{ref}/\hbar)^{-1/2}=\SI{2.0e-12}{\radian}$ and $L_\text{ref}\equiv(\hbar I \Omega_\text{ref}/2)^{1/2}=\SI{2.6e-23}{\joule\cdot\second}$, chosen to render $\theta$ and $L$ variances the same in magnitude. \textbf{c.} Spectral noise densities for causally estimated angular position/momentum trajectories, and for deviations between causal and smoothed estimates, encoding conditional noise information. The residuals between the measured and estimated angular position are a proxy for the in-situ detection noise. The effective detection noise level indicates the closest white-imprecision-noise model replicating observations. For general properties of the expected model spectra, see SI. \textbf{d.} Cumulative contributions to the covariance elements of the observed deviation $V_\leftrightarrow$, shown as a function of the upper frequency integration cutoff: Most of the contribution accumulates near frequencies centered around $\Omega_\text{ref}/(2\pi)=6\,\text{Hz}$.}\label{fig:State}
\end{figure}

In order to extract properties of the measurement conditioned states, say at $t_0$, from the recorded angular time traces, we reconstruct phase space trajectories of the rotor's conditional mean, initially through \emph{causal estimation} of the conditioned state using only past data up to time $t_0$, then through \emph{smoothed estimation}, gathering future data after $t_0$ about the conditional state at $t_0$, refining our knowledge. These estimation protocols are formally the Kalman-Bucy~\cite{kalman1961new,wiseman2009quantum} and the RTS smoothing filters~\cite{rauch1965maximum,laverick2021quantum}, respectively. Given the dynamical model of the system, these time-domain filters optimally minimize the theoretical phase-space uncertainty around the estimated conditional means at $t_0$. The corresponding estimated motions of our rotor (Fig.~\ref{fig:State}a) yield unconditional noise spectra (Fig.~\ref{fig:State}c) that follow the expected frequency scaling for angular position and momentum, respectively, with the corresponding PSDs $S_{\theta\theta}=S_{\tau\tau}|\chi|^2$ and $S_{LL}=(I\omega)^2S_{\theta\theta}$, where $\chi(\omega)=-1/(I\omega^2)$ is the susceptibility of the free rotor.

For these estimations, the motion of our rotor is modeled by a conditional Markovian quantum master equation with augmented degrees of freedom to reproduce the time-correlated structural thermal noise. In the Gaussian state limit, the dynamics are captured by the conditional quantum expectations of the angular position and momentum through the vector $R=\{\langle\theta\rangle,\langle L\rangle\}^{\mathsf T}$, and the $2\times2$ symmetric conditional covariance matrix $V$ encoding the Gaussian phase-space uncertainty, with the elements $V_{\theta\theta}$, $V_{\theta L}$ and $V_{LL}$, and a corresponding state purity $\mathcal{P}=1/\sqrt{\det V}$. Whereas $R$ obeys a linear stochastic differential equation, $V$ obeys a nonlinear deterministic differential equation. The relevant corresponding quantities for the smoothed states are the vector $R_s$ and the smoothed covariance matrix $V_s$. Note that, because of the augmentation, in practice we use a four-element state vector. The full machinery of this analysis is detailed in SI.

The fact that smoothing invariably yields a more precise estimate of a conditional state---an effect that is particularly pronounced for free-particle dynamics---provides a conceptually straightforward way to determine the conditional uncertainties experimentally. With smoothed states taken as a phase-space reference, an estimate of the covariance matrix $V$ of states prepared via causal estimation is expected to approximately coincide with the covariance matrix $V_\leftrightarrow=\mathrm{Var}(R_\leftrightarrow)$ of the experimentally observed deviation \mbox{$R_{\leftrightarrow}=R-R_s$} between the causal and smoothed conditional means. In practice, this simple association leads to a slight underestimation due to correlations between the causal and smoothed estimates. The mathematically accurate estimator determined from the fundamental orthogonality principle for conditional expectations (see SI) can be written as
\begin{equation}
V^\text{(est)} = V_\leftrightarrow+V_s. \label{Eq:estimator}
\end{equation}
Here, the (theoretical) smoothed covariance matrix $V_s$ acts as a correction to the experimentally extracted quantity $V_\leftrightarrow$, producing changes of only about $10$–$20\%$ in the individual estimated matrix elements for the present analysis. A principal utility of Eq.~\ref{Eq:estimator} lies in its ability to determine the model parameters that provide the closest agreement with the observed data, as the equation must be solved self-consistently: $V_\leftrightarrow$ implicitly depends on the model parameter-dependent theoretical values of both $V$ and $V_s$ supplied to the causal and smoothed estimation algorithms. For a successful estimate, the input $V$ must match the obtained $V^\text{(est)}$.

Because the thermal noise level is well established for our setup (Fig.~\ref{fig:Experiment}b), the only free parameter in the estimation model is the effective (white) detection-imprecision noise level. We determine this single parameter by solving Eq.~\ref{Eq:estimator} self-consistently and minimizing the error matrix \mbox{$\Delta = V^\text{(est)}-V$}; specifically, we minimize the maximum fractional covariance disagreement over any phase-space direction, obtaining a largest residual discrepancy of $1.5\%$ at the optimum. This procedure yields an effective imprecision noise level of \mbox{$39\pm 1\,\text{prad/}\sqrt{\text{Hz}}$}, with the 1-$\sigma$ uncertainty due to the finite data size---see SI for optimization and statistical error analysis. The imprecision level obtained is consistent with the mean in-situ detection noise implied by the spectral residuals between the measured and estimated trajectories (Fig.~\ref{fig:State}c) across the \qtyrange[range-units=single,range-phrase=-]{3}{12}{\hertz} band, where the covariance estimator picks up most of its magnitude (Fig.~\ref{fig:State}d). We can also construct an estimator for the smoothed covariance matrix by turning around Eq.~\ref{Eq:estimator}, obtaining  \mbox{$V^\text{(est)}_s = V-V_\leftrightarrow$}, with the same model parameters optimizing self consistency. We visualize the estimated conditional phase-space distributions (Fig.~\ref{fig:State}b) by streaming the transformed deviation vectors $\tilde{R}_\leftrightarrow=M\cdot R_\leftrightarrow$, where $M$ is chosen to produce $V^\text{(est)}=\mathrm{Var}(\tilde{R}_\leftrightarrow)$ or $V^\text{(est)}_s=\mathrm{Var}(\tilde{R}_\leftrightarrow)$.

The results imply \mbox{$\mathcal{P}=(1.45\pm 0.04)\times10^{-4}$} and \mbox{$\mathcal{P}_s=(5.8\pm 0.2)\times10^{-4}$} for the purities of causally prepared and smoothed conditioned states, respectively. The latter level of purity matches that of a (sub-\SI{10}{\hertz}) oscillator cooled to just $860$ mean thermal quanta. By inserting $E=V_{LL}/2I$ into Eq.~\ref{Eq:temperature}, valid for both conditioning approaches, we find a temperature of $1.20\pm 0.02\,\mu\text{K}$ for causally prepared states and $140\pm 20\,\text{nK}$ for smoothed conditional states. The achieved purities/temperatures would not have been possible with conventional feedback cooling due to the unavoidable imprinting of high-frequency measurement-imprecision noise on angular momentum~\cite{komori2022quantum}, and smoothed states would not have yielded such large information gain if the states already had near-unity purity~\cite{laverick2019quantum}. The inferred sub-$\mu\text{K}$ effective temperatures bring optomechanical systems into the ultracold regime, previously associated mainly with evaporatively cooled atomic gases.

\begin{figure}[h]
\centering
\includegraphics[width=0.5\textwidth]{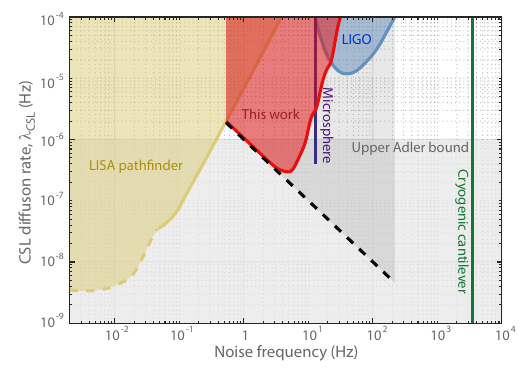}
\caption{Low frequency landscape for continuous spontaneous localization models. Red shaded region is the exclusion in this work, penetrating below the upper Adler bound~\cite{adler2007lower}. The light shaded region above the black dashed-line is the foreseeable future exclusion by using an optical cavity instead of an optical lever. Regions excluded by other low frequency experiments are also shown: Advanced LIGO and LISA pathfinder space mission~\cite{helou2017lisa,armano2018lisa} as well as levitated microsphere~\cite{zheng2020microsphere} and cryogenic cantilever ~\cite{vinante2020cantilever} experiments.}\label{fig:CSL}
\end{figure}

\section{Sensing frontier and outlook}\label{sec:Sensing}
The free-particle regime established here offers immediate utility for tests of fundamental physics by combining exceptional isolation and access to low frequencies with the sensitivity required to resolve residual motional fluctuations. As a first demonstration of this utility, we use the directly measured fluctuations to constrain proposed modifications of quantum mechanics, excluding new parameter space extending into the Adler range~\cite{adler2007lower} of continuous spontaneous localization (CSL) models at low frequencies. We then exploit the utility of the low-entropy smoothed states to assess the prospects for gravity-mediated entanglement and highlight the significant improvement over any previously demonstrated system targeting quantum control. In this connection, we point out that whereas causal conditioning provides the instantaneous state available for applications such as feedback control, the smoothed state---closely related to the past quantum state~\cite{gammelmark2013past}---has direct operational significance, enabling optimal estimation of unknown forces~\cite{iwasawa2013quantum} and retrospective characterization of quantum entanglement~\cite{bao2020spin}.

%Whereas causal conditioning provides the instantaneous state available for applications such as feedback control, the smoothed state---closely related to the past quantum state~\cite{gammelmark2013past}---has direct operational significance, enabling optimal estimation of unknown forces~\cite{iwasawa2013quantum} and retrospective characterization of quantum entanglement~\cite{bao2020spin}. The free-particle regime established here offers immediate utility for tests of fundamental physics by combining exceptional isolation and access to low frequencies with the sensitivity required to resolve residual motional fluctuations. As a first demonstration of this utility, we use the directly measured fluctuations to constrain continuous spontaneous localization (CSL) models, which propose modifications of quantum mechanics, excluding new parameter space extending into the Adler range~\cite{adler2007lower} at low frequencies. We then exploit the additional information obtained through smoothing to assess the prospects for gravity-mediated entanglement, highlighting the significant improvement over any previously demonstrated system targeting quantum control.

CSL introduces wavefunction collapse and its accompanying momentum diffusion as a mechanism for the emergence of classical behavior. While stringent constraints on model parameters have been obtained from x-ray emission experiments~\cite{carlesso2022present,aprile2026challenging}, the bounds derived from high-frequency phenomena become ineffective for models with temporally correlated noise~\cite{torovs2017colored}. Low-frequency experiments therefore provide a complementary testing ground. Evolution under CSL predicts an excess noise for a rigid rotor with torque PSD $S_{\tau\tau}^{CSL}(\omega)=\lambda_\text{CSL}8\pi\hbar^2(r_\text{CSL}/m_u)^2(\rho I/d)$~\cite{diosi2015testing,komori2020attonewton}. Here, $\rho = \SI{2.2}{\gram\per\cm^3}$ is the rotor's mass density, $d=\SI{0.5}{\milli\meter}$ is its thickness, and $m_u$ is the atomic mass unit. The collapse length scale $r_\text{CSL}$ is conventionally chosen as \SI{100}{\nano\meter}~\cite{adler2007lower}, a regime in which experimental tests have proven more difficult. The measured angular noise floor achieved in the present experiment translates into an exclusion of the diffusion rate $\lambda_\text{CSL}$ illustrated in Fig.~\ref{fig:CSL}, excluding regions complementary to those ruled out by the LISA Pathfinder mission.

More broadly, the operating regime established here is directly relevant to experiments at the interface of quantum mechanics and gravity, for which the candidate systems can be benchmarked~\cite{agafonova2026one} with a quantum-gravity figure-of-merit $\eta_{QG}$, where gravity-mediated entanglement would require $\eta_{QG}\geq1$ alongside a sufficiently high achievable single particle purity. In SI, we extend these criteria to a continuously monitored free particle, building on earlier work on harmonic oscillators~\cite{miki2023generating}, finding that a purity greater than 0.4 is not needed to generate entanglement. Despite relying solely on optical-lever readout, the current setup implies $\eta_{QG} = 0.8\times10^{-6}$. While underscoring the considerable challenge of achieving gravitational entanglement, this value means that, for two identical rotors, the achievable ratio of entanglement generation to decoherence rate would be nearly two orders of magnitude higher than that could be inferred for any previously demonstrated system~\cite{agafonova2026one}. For our platform, reaching a purity of 0.4 is within reach by incorporating a cavity-enhanced torsional readout~\cite{agafonova2024zigzag} into the current system, using a modest cavity finesse of one thousand. Such an advance alone would substantially extend the reach of fundamental tests of quantum physics and gravitation~\cite{bose2025massive}.

%\section{Notes}
%\begin{itemize}
%    \item 
%\end{itemize}

\backmatter

\bmhead{Supplementary information}

This article contains a supplementary information file detailing the experimental setup, its characterization, and the theoretical analysis tools.

\bmhead{Acknowledgements}

We thank Nobuyuki Matsumoto for fruitful exchanges on the topic of mechanical state verification, and Ugur Sezer for assistance in microfabrication of the rotor substrates.

\section*{Declarations}

\begin{itemize}
\item Funding: This work was supported by Institute of Science and Technology
Austria, the European Research Council under Grant No. 101087907 (ERC CoG
QuHAMP), and Austrian Science Fund (FWF) [10.55776/COE1] (Cluster of Excellence quantA).
\item Conflict of interest/Competing interests: Authors declare no competing interests.
\item Ethics approval and consent to participate: Not applicable.
\item Consent for publication: Authors provide consent for publication.
\item Data availability: Data available upon reasonable request.
\item Materials availability: Not applicable.
\item Code availability: Data analysis code is available upon reasonable request. The implemented analysis is readily detailed in the SI.
\item Author contribution: UM, TS and OH conceived and built the experiment, collected the data, and analyzed the results. GVG, FGL, MM and OH developed theoretical aspects and analysis methods. All authors contributed to the preparation of the manuscript.
\end{itemize}

\bibliography{free-particle}% common bib file
%% if required, the content of .bbl file can be included here once bbl is generated
%%\input sn-article.bbl

\includepdf[pages=-]{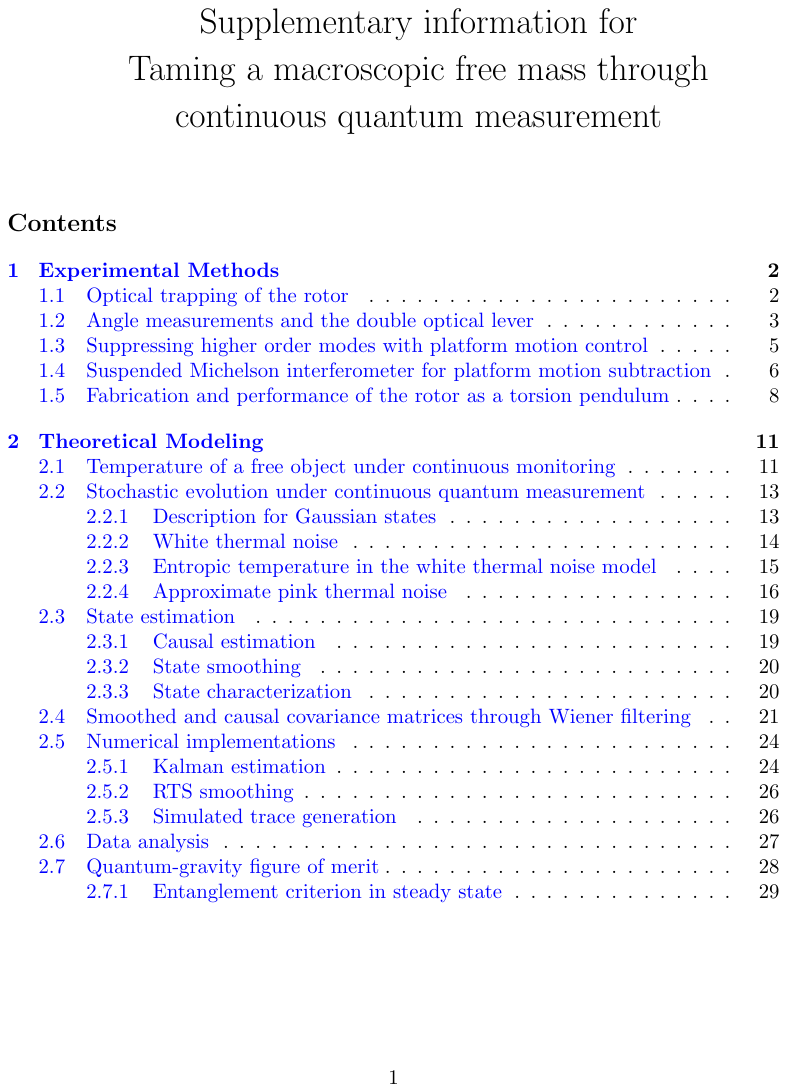}

\end{document}